# Thermoelectric Potential of NaVAs Half-Heusler Alloy: Insights from Ab-initio Calculations*

Rajinder **Singh**[a], Shyam Lal **Gupta**[b], Sumit **Kumar**[c], Lalit **Abhilashi**[a], Diwaker[d,*] and Ashwani **Kumar**[a,**]

[a]School of Basic Sciences, Abhilashi University Mandi, Mandi, 175045, H P , INDIA

[b]Exploring Physics for Interdisciplinary Science and Technology(EPIST) lab, Harish-Chandra Research Institute, Prayagraj, Allahabad, 211019, U P , INDIA

[c]Department of Physics, Government College, Una, 174303, H P , INDIA

[d]Department of Physics, S. C. V. B. Government College, Palampur, Kangra, 176061, H P , INDIA

## ABSTRACT

This work presents a comprehensive investigation of the HH alloy NaVAs using first-principles methods, emphasizing its potential applications in various fields, including spintronics, thermoelectrics, and optoelectronics. We utilized density functional theory (DFT) within the full-potential linearized augmented plane wave (FLAPW) framework. Structural optimizations indicate that the alloy stabilizes in the ferromagnetic (γ) phase. Both mechanical and dynamical stability have been confirmed through analysis of elastic constants and phonon dispersion. Our calculations of the electronic band structure and density of states (DOS) reveal that NaVAs exhibits half-metallic behavior, with a spin-polarized band gap of 2.77 eV in the minority spin channel. The magnetic moment aligns with the Slater-Pauling (SP) rule, demonstrating robust ferromagnetism. Mechanical analysis shows that the material is brittle in nature. The thermodynamic analysis highlights the alloy's resilience, supported by consistent trends in entropy, heat capacity, and Debye temperature. Its optical response indicates strong absorption in the visible and ultraviolet (UV) regions, along with pronounced dielectric and plasmonic features, suggesting potential for applications in optoelectronics and reflective coatings. Furthermore, evaluations of the transport properties show high Seebeck coefficients and a significantly tunable figure of merit (ZT). ZT values approach 1.0 across the temperature range of 600-1500 K, demonstrating excellent thermoelectric characteristics. Overall, these findings position NaVAs as a promising multifunctional material suitable for advanced technological applications in green energy area.

## 1. Introduction

Every year, the need for energy sources and storage solutions grows on a global scale. At the same time, choosing energy sources that are pollution-free, renewable, and ecologically friendly is highly valued. Thermoelectric power generating stands out among the other possibilities as a wise decision for the modern period. Thermoelectric materials can be used to transform the waste heat produced by numerous daily activities into electrical power [1, 2, 3, 4, 5, 6, 7, 8, 9, 10, 11, 12]. Thermoelectric devices are used in many different technological fields besides power generating. They can be utilized in wireless microelectronic devices and sensors, as well as in medical wearable health monitoring equipment. Thermoelectric materials are employed in the aircraft industry to provide electricity in harsh environments. The history of the thermoelectric effect's discovery is intriguing. Alessandro Volta made the initial observation of it in the early 1800s while studying how muscles contract when exposed to temperature changes. He came to the conclusion that an electromotive force (emf) is produced between the connections of two distinct metal conductors by a temperature differential. Johann Seebeck, a German physicist, noticed thermocouple-induced variations in a magnetic compass some 20 years later. He misunderstood this event, thinking that a magnetic field was created by the temperature differential. But later on, his investigations helped to validate the thermoelectric effect, which Oersted subsequently investigated and named "thermoelectric." There are now many uses for thermoelectric devices, and research is being done to find novel materials and creative ways to improve their functionality [13, 14, 15, 16, 17, 18]. The figure of merit (ZT) is used to evaluate a thermoelectric material's effectiveness, which is influenced by multiple factors. The following formula is used to determine the figure of merit:

$$ZT = \frac{S^2 \sigma T}{\kappa} \qquad (1)$$

Here, S stands for the Seebeck coefficient, $\sigma$ for the electrical conductivity, and $\kappa$ for the thermal conductivity of the chosen thermoelectric material in the equation above. A thermoelectric material's figure of merit indicates its performance; larger the figure of merit, better will be material's performance. High electrical conductivity, low thermal conductivity, and a high Seebeck coefficient are desirable properties for thermoelectric materials. Among the well-known thermoelectric materials on the market are

---

*NaVAs half-heusler alloy

*Corresponding author

**Corresponding author

diwakerphysics@gmail.com ( Diwaker); ashwani.bits05@gmail.com (A. Kumar)

ORCID(s): 0000-0002-4155-7417 ( Diwaker); 0000-0002-1078-908X (A. Kumar)

silicon-germanium alloys, lead telluride, and bismuth telluride. Skutterudites, half-Heusler (HH) alloys, clathrates, oxy-selenides, chalcogenides, organic hybrid materials, and nanostructures are among several other thermoelectric materials that are presently the subject of investigation. We now focus on half-Heusler compounds, whose multifunctional qualities have been the subject of much research. Excellent thermoelectric materials with high thermoelectric efficiency (ZT) throughout a wide range of operating temperatures are half-Heusler (HH) alloys [19, 20, 21, 22]. To improve their thermoelectric performance, HH materials have been investigated using a variety of techniques, including doping, substitution, vacancies, defect engineering, and dimension reduction. Two d-block elements and one p-block element are the usual components of HHs. It was found that the compound that resulted from the alloying of nonmagnetic elements in HH alloys had magnetic characteristics. It has been noted that HH alloys display half-metallic (HM) characteristics. This indicates that the material works like a semiconductor in one spin orientation and like a metal in another. The particular arrangement of atoms inside the materials, their interactions with one another, and the total quantity of valence electrons give rise to these special qualities. For spintronic devices such as magnetic random access memory (MRAM) devices, spin injectors, and spin valves, these alloys' high magnetoresistance (HM) feature is highly valued. Many V-based Heusler alloys have not yet been fully investigated, and while vanadium-based Heusler alloys have been documented, there are currently very few experimental research available. In this study, we identify Heusler alloys based on vanadium, namely NaVAs, in order to learn more about their thermoelectric and spintronic characteristics.

## 2. Computational Details

By using the Kohn–Sham (KS) equation, which reduces the many-body electron problem to a single-body electron problem, DFT is an effective technique for tackling this problem. To solve the self-consistent single-electron problem, we used the all-electron code package WIEN2k, which uses full potential linearized augmented plane wave (FP-LAPW) techniques. We were able to compute the NaVAs HH from first principles as a result. We used the Birch–Murnaghan equation to fit the energy versus volume data in order to get the optimal lattice parameters. For our self-consistent ground state energy calculations, we used the flexible pseudopotentials and exchange-correlation (XC) functionals from WIEN2K [23, 24]. The generalized gradient approximation (GGA) is the most widely utilized XC functional. In addition to the results from the GGA approach, this study also provides results from the TB-mBJ method, which show enhanced accuracy in the electrical density of states (DOS) and band structure. [25]. The Phonopy algorithm with the finite displacement approach was used to calculate phonon dispersion. The mechanical stability of the HH alloy NaVAs was determined using the IRelast package that comes with WIEN2k. To distinguish between valence and core electrons, a constant energy value of -6.0 Ry was used. Brillouin zone sampling was conducted using a 6x6x6 K-point mesh in accordance with the Monkhorst–Pack scheme. For every atom involved, the muffin tin (RMT) radii were RMT [Na] = 2.43, RMT [V] = 2.50, and RMT [As] = 2.50. The number of generated plane waves was limited by setting the cut-off value RMT × KMAX at 7. Using the quasi-harmonic approximation, the thermodynamic properties of our materials were investigated using the GIBBS2 software within the melting point temperature range. The BoltzTrap2 code, which assumes a constant relaxation time and works under the rigid band approximation, was used to compute the transport properties. As we raised the temperature from 600 to 1500 K, we computed the Seebeck coefficient, electrical conductivity, and electronic thermal conductivity.

## 3. Results and Discussion

The computational results of the different physical properties of the NaVAs HH alloy will be covered in this section.

### 3.1. Structure and stability

Our first goal is to investigate the NaVAs HH alloy's structural stability. The face-centered cubic system, space group number 216 (F-43m), is where this HH alloy crystallizes [26, 27]. Based on atomic poistions, the various structures of this HH alloy can be divided into alpha ($\alpha$), beta ($\beta$), and ($\gamma$) structures. The $\alpha$-XYZ structure has the following Wyckoff positions: Z (0, 0, 0), Y (0.25, 0.25, 0.25), and X (0.5, 0.5, 0.5). The Wyckoff locations for the $\beta$-XYZ structure are Z (0, 0, 0), Y (0.5, 0.5, 0.5), and X (0.25, 0.25, 0.25). The Wyckoff locations in the $\gamma$-XYZ crystal structure are Z (0.5, 0.5, 0.5), Y (0.25, 0.25, 0.25), and X (0, 0, 0). Our calculations revealed that the ($\gamma$)-XYZ structure is the only one reported here and has the lowest energy. By examining the volume-energy curve and fitting it with the Birch–Murnaghan equation of state, which is shown below, the ideal lattice parameters for this alloy were identified.

$$E_{tot}(V) = E_0(V) + \frac{B_0 V}{B_0(B_0 - 1)}\left[B\left(1 - \frac{V_0}{V}\right) + \left(\frac{V_0}{V}\right)^{B'_0} - 1\right] \tag{2}$$

The minimum energy values and other parameters for the NaVAs HH alloy's optimization for both ferromagnetic and non-magnetic configurations are given in Table 1. The Birch–Murnaghan equation, which is shown in fig. 1 for the alloy's ($\gamma$) structure, is used to fit the optimization curves. Figure 1 displays the ($\alpha$) phase's visual representation, which was produced with Xcrysden software. It is evident that the NaVAs HH alloy has the lowest energy configuration in the ferromagnetic order. The Birch–Murnaghan fit of the volume versus energy curve for the HHs NaVAs yielded optimal lattice parameters of 6.2804 Å. The formation energy ($E_{form}$) in Ryd is expressed as follows:

$$E_{form} = E_{NaVAs} - (E_{Na} + E_V + E_{As}) \tag{3}$$

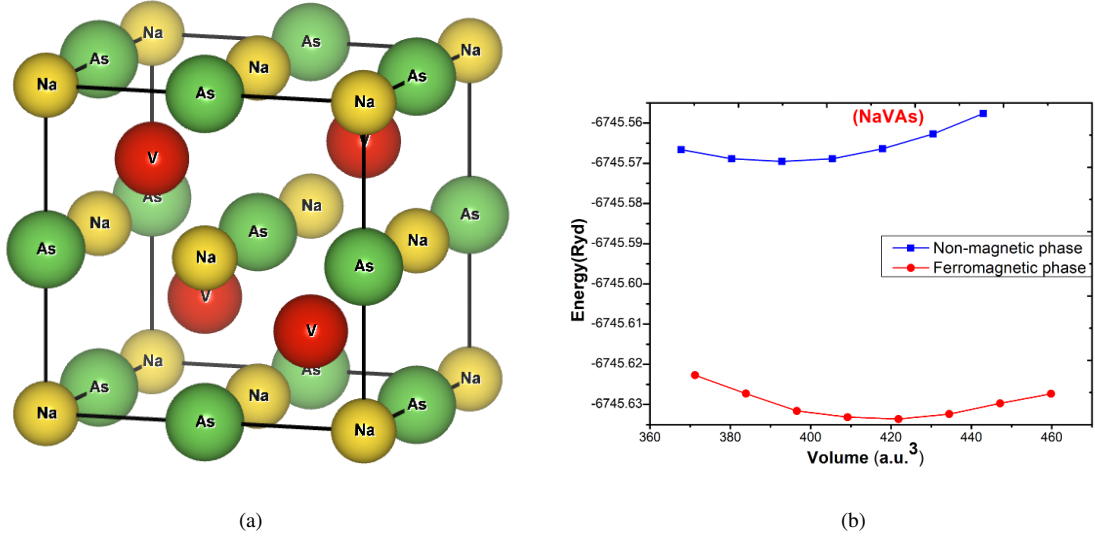

**Figure 1:** [a] Optimized unit cell and [b] Energy versus Volume curve of NaVAs HH alloy

**Table 1**
Optimized structural and other parameters of NaVAs HH alloy.

| NaVAs | a(Å) | $V_o$(GPa) | $E_o$(Ryd) |
|-------|------|-----------|-----------|
| SP | 6.2804 | 417.9323 | -6745.633391 |
| NM | 6.1423 | 390.9604 | -6745.569683 |

The potential energy minimum of the HHs NaVAs is denoted by $E_{NaVAs}$ in this equation. The minimum energy of the individual atoms Na, V, and As is indicated by the variables $E_{Na}$, $E_V$, and $E_{As}$, respectively. The formation energy $E_{form}$ from the OQMD database was determined for NaVAs HH alloy and is found to be -0.508eV/Atom. This -ve value of formation energy acertain the possibility of experimental synthesis of NaVAs HH alloy. Also, the stability values are 0.080 eV per atom for the NaVAs Alloy [28, 29].

## 3.2. Mechanical stability
Using the IRelast package in Wien2k, we computed the stiffness matrix (s) in accordance with equation (4) to examine the mechanical stability of our NaVAs HH alloy. Only three different values, which represent the three elastic constants, are present in the stiffness matrix of cubic crystal systems: $C_{11}$, $C_{12}$, and $C_{44}$ [30, 31, 32]. As we get further into this section, we will talk about the range of mechanical properties that these constants give for each HH alloy.

$$\sigma = c_{ij} \tag{4}$$

Equation (5-9) provides the Born and Huang stability criterion, which can be used to verify the mechanical stability of a cubic crystal system. [33]. All our HH alloys meet these conditions, thus confirming their mechanical stability.

$$c_{11} - c_{12} > 0 \tag{5}$$
$$c_{11} > 0 \tag{6}$$
$$c_{11} + 2c_{12} > 0 \tag{7}$$

$$c_{12} < B < c_{11} \tag{8}$$
$$c_{44} > 0 \tag{9}$$

Our HH NaVAs' bulk modulus (B) was determined by applying equation (10). This modulus gives information about a material's resistance to volumetric forces. A high bulk modulus means that when pressure is applied, the crystal deforms less. According to our calculations, HH NaVAs have a considerable bulk modulus of 56.257 GPa, indicating their resistance to deformation.

$$B = \frac{(c_{11} + 2c_{12})}{3} \tag{10}$$

The resistance of a material to surface forces is measured by its shear modulus. Greater resistance to shear pressures is indicated by a bigger shear modulus. Equation (11), the formula for calculating the shear modulus, was used

$$G = \frac{(c_{11} + c_{12} + c_{44})}{5} \tag{11}$$

and 145.907 GPa is the value that is computed. Equation (12) can be used to determine Young's modulus from the bulk and shear modulus as

$$E = \frac{9BG}{3B + G} \tag{12}$$

and is found to be 234.762 GPa. The tensile strength of materials is shown by their Young's modulus. Vickers' hardness formula (eqn (13)) was used to measure the hardness of our materials [34]. NaVAs was discovered to have the highest hardness value, measuring 81.811 GPa.

**Table 2**
List of important mechanical parameters of NaVAs HH alloys

| Parameter | NaVAs |
|---|---|
| Elastic constant $C_{11}$ (in GPa) | 134.478 |
| Elastic constant $C_{12}$ (in GPa) | 17.146 |
| Elastic constant $C_{44}$ (in GPa) | 204.067 |
| Bulk Modulus B (in GPa) | 56.257 |
| Young's Modulus E (in GPa) | 234.762 |
| Shear Modulus G (in GPa) | 145.907 |
| Pugh's ratio ($\frac{B}{G}$) (in GPa) | 0.453 |
| Anisotropy factor A (in GPa) | 134.478 |
| Lame's $1^{st}$ constant $\lambda$ (in GPa) | -26.537 |
| Lame's $2^{nd}$ constant $\mu$ (in GPa) | 124.191 |
| Kleinman parameter $\zeta$ (in GPa) | 0.299 |
| Vicker's hardness $H_v$ (in GPa) | 81.811 |

$$H_v = \left[\left(\frac{G}{B}\right)^2 \times G\right]^{0.585} - 3 \qquad (13)$$

Understanding the bending and stretching dynamics of atomic bonds is crucial for comprehending a variety of materials' macroscopic mechanical and chemical characteristics. The degree of bond bending and stretching in materials is measured by the Kleinman parameter $\zeta$ (in GPa), which is provided by equation (14). Bond bending reaches its maximum for materials with a Kleinman parameter of $\zeta$ (in GPa) = 1, whereas bond stretching reaches its maximum for materials with z = 0. With a Kleinman value $\zeta$ (in GPa) of roughly 0.2 for the investigated NaVAs HH alloy, these materials mostly show a bond-bending character..

$$\zeta = \frac{c_{11} + 8c_{12}}{7c_{11} + 2c_{12}} \qquad (14)$$

The compressibility and shear stiffness of materials are indicated by Lame's first constant ($\lambda$) and second constant ($\mu$), respectively. Equation (15-16) illustrates how these constants are determined using the Poisson ratio and Young's modulus. This leads us to the conclusion that NaVAs have superior shear strain resistance.

$$\lambda = \frac{E_v}{(1-v)(1+2v)} \qquad (15)$$

$$\mu = \frac{E}{2(1+v)} \qquad (16)$$

The calculated values for elastic constants as reported in Table 2. comply to the required Born's criteria of mechanical stability represented by eqn. (5-9). Further, Pugh's ratio ($\frac{B}{G}$) of the bulk and shear modulus, which is less than 1.75 indicate the brittle nature of NaVAs HH alloy.

### 3.3. Electronic properties

Understanding the connection between a material's structure and properties is based on its band structure. It offers useful information that enables us to categorize materials according to how their bands are arranged in respect to the Fermi level. The spin-polarized (up and down) band structure for our HH alloys, as determined by both GGA and mBJ functionals, is displayed in Figure 2. While the Fermi energy level ($E_F$) is denoted by horizontal dotted lines and named as such, the spin-up and spin-down configurations are designated with up and down. The band structure shape close to the Fermi level does not significantly change for high-harmonic NaVAs. Nevertheless, both GGA and TB-mBJ computations show a discernible shift in the valence band maximum (VBM) and the conduction band minimum (CBM). An indirect band gap is shown by the CBM and VBM not lying at the $\Gamma$ point. HH NaVAs is an indirect band gap material as a result. The band structure plots displayed in Figure 2 reveal the band gap. Similar band topologies for HH NaVSb are predicted by the GGA and TB-mBJ approaches; however, it is crucial to remember that the band gap values are increased in the TB-mBJ computations [25]. When forecasting the band gap for solids, the semi-localized GGA exchange correction is known to perform badly. The TB-mBJ functional, on the other hand, uses an adjusted screening parameter that lowers the band gap prediction error. Equation (18) shows that our HH alloys are half-metallic due to 100% spin polarization at the Fermi energy level.

$$P = \frac{N_\uparrow(E_F) - N_\downarrow(E_F)}{N_\uparrow(E_F) + N_\downarrow(E_F)} \qquad (17)$$

Due to the energy levels overlapping the Fermi level, our alloys display metallic characteristics in the spin-up position. On the other hand, the noticeable band gap close to the Fermi level makes them semiconducting in the spin-down state. Comparing the GGA and TB-mBJ exchange-correlation (XC) functionals, Figure 3 displays the total density of states (DOS) for our HH NaVAs' spin-down and spin-up configurations. It is evident from fig. 3 (2nd row) that both GGA and TB-mBJ calculations do not reveal any energy states at the Fermi level for NaVAs. Figure (3) indicates that the d-block element vanadium (V) close to the Fermi level is the main source of energy states. The interactions between various atoms within the unit cell, especially their orbital interactions, produce energy band gaps. The hybridized orbitals that are produced as a result of these interactions are crucial for the development of the band gaps. It is evident from Figure 4 that As atoms contribute more in the valence band while V atoms contribute more in the conduction band. The band gap in our materials is caused by these hybridizations. The orientation of electron spins and a material's spin magnetic moment (MM) influence its magnetic characteristics. The Slater-Pauling rule (SPR) can be used to determine the total magnetic moment (MT) of a half-metallic (HH) material containing alkali metals, as indicated by the following equation:

$$M_T = Z_t - 8 \qquad (18)$$

The total number of valence electrons is denoted by $Z_t$ in the equation above. Since our HH NaVAs have eleven valence electrons, we can determine $M_T = 3\mu B$ based on the SPR. We calculated the overall and individual atomic magnetic moments using our self-consistent field. Table 3 displays these values. The SPR is well satisfied by the computed MM. Our HH NaVAs clearly show ferromagnetic activity when the individual magnetic moments are examined. The finding that every atom has a positive magnetic moment leads to this conclusion. Consequently, it is clear that the system as a whole exhibits ferromagnetic behavior because of the preponderance of positive magnetic moments that each atom contributes. The total magnetic moment's integral value shows that our material has half-metallic characteristics. Half metals behave like semiconductors for one spin orientation and like metals for the other. There are numerous uses for these materials in spintronic devices.

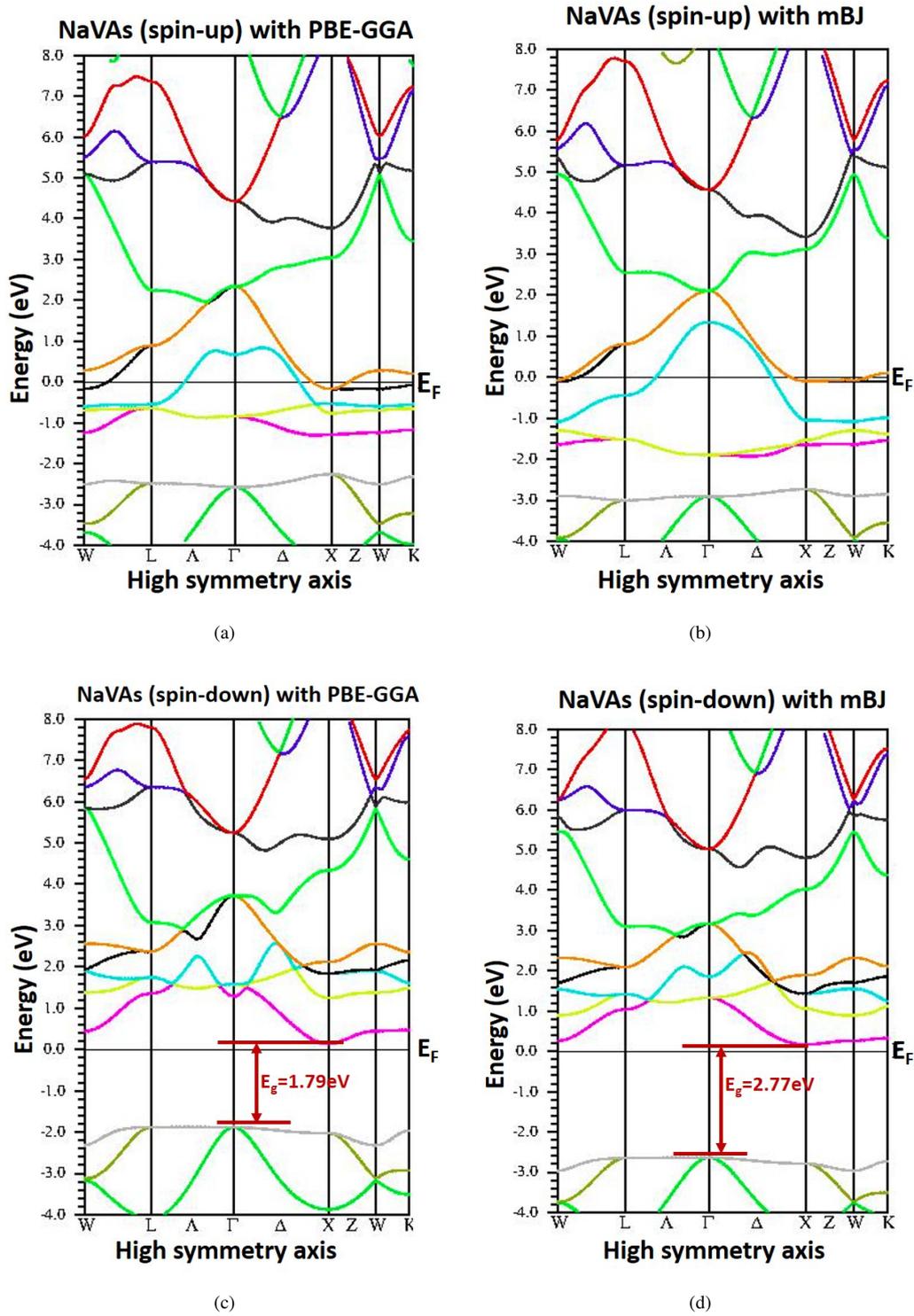

**Figure 2:** Spin-up and spin-down electronic band structure plot of NaVAs HH alloy using GGA and mBJ approach

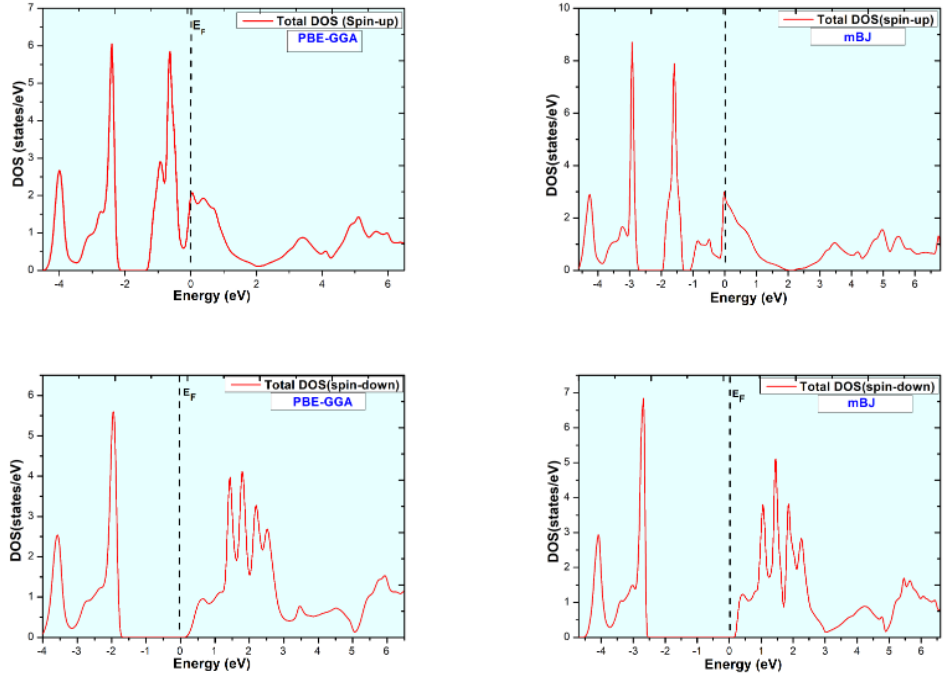

**Figure 3:** Total DOS calculated for HH NaVAs for spin-dn and spin-up configurations using GGA and TB-mBJ exchange–correlation functionals.

**Table 3**
Value of Magnetic moment for NaVAs per unit cell

| | |
|---|---|
| Total $M_T$ ($\mu_B$) | 2.99992 $\mu_B$ |
| MM of Na ($\mu_B$) | 0.32629 $\mu_B$ |
| MM of V ($\mu_B$) | 0.07191 $\mu_B$ |
| MM of As ($\mu_B$) | 2.73367 $\mu_B$ |
| Interstitial MM ($\mu_B$) | -0.13195 $\mu_B$ |

## 4. Transport Properties

The capacity of thermoelectric materials to transform heat energy into electrical energy and vice versa is well known. We computed the Seebeck coefficient (S), electronic thermal conductivity per relaxation time ($\frac{\kappa_e}{\tau}$), electrical conductivity per relaxation time ($\frac{\sigma}{\tau}$), and the figure of merit (ZT) in order to evaluate the thermoelectric performance of the NaVAs alloy. The BoltzTrap code's implementation of the Boltzmann transport theory was used for these computations [35]. The Constant Relaxation Time Approximation (CRTA) serves as the foundation for these computations. The dimensionless figure of merit, or ZT, is a measure of thermoelectric material efficiency and is defined by the following equation:

$$ZT = \frac{S^2 \sigma T}{\kappa} \tag{19}$$

In the above equation, T signifies absolute temperature, and $\kappa$ refers to total thermal conductivity, which includes both electronic ($\kappa_e$) and lattice ($\kappa_l$) contributions and is given as

$$\kappa = \kappa_e + \kappa_l \tag{20}$$

### 4.1. Seebeck Coefficient

The amount of thermoelectric voltage produced in response to a temperature gradient across a material is measured by the Seebeck coefficient (S), also known as thermopower. Because it has a direct impact on the power factor and the overall figure of merit (ZT), this coefficient is an important parameter. It is given by

$$S = \frac{1}{|e|T} \cdot \frac{\int (\epsilon - E_F)\sigma(\epsilon)(\frac{-\partial f}{\partial \epsilon})d\epsilon}{\int \sigma(\epsilon)(\frac{-\partial f}{\partial \epsilon})d\epsilon} \tag{21}$$

The elementary charge is denoted by e, the temperature by T, the energy by $\epsilon$, the fermi energy by $\mu$, the energy-dependent conductivity by $\sigma(\epsilon)$, and the Fermi-Dirac distribution function by f in the equation as above. Figure 5a illustrates the calculated Seebeck coefficient as a function of $(\mu - E_F)$ in [Ha] where $\mu$ represents the individual energies for NaVAs alloy at different temperatures ranging from 600 K to 1500 K. The plot shows two dominant peaks: one positive and one negative, located on left side of the Fermi level. The maximum magnitude of the Seebeck coefficient (S) is observed in the range -0.1 to -0.05 [Ha] away from the Fermi level ($E_F$). The positive peak on the left indicates p-type behavior, while the negative peak signifies n-type conduction. As the temperature rises from 600 K to 1500 K, the Seebeck peaks decreases while their width increase. This change is attributed to the enhanced thermal excitation of charge carriers, which smooths the Fermi-Dirac distribution and reduces the energy dependence of the carrier population [36]. A sharp transition in the sign of the Seebeck coefficient occurs at $(\mu - E_F = -0.09)$ [Ha]. This indicates the penetration of fermi level into the conduction band which is consistent with the observed band structure and electronic DOS shown in fig. 2 and fig. 3.

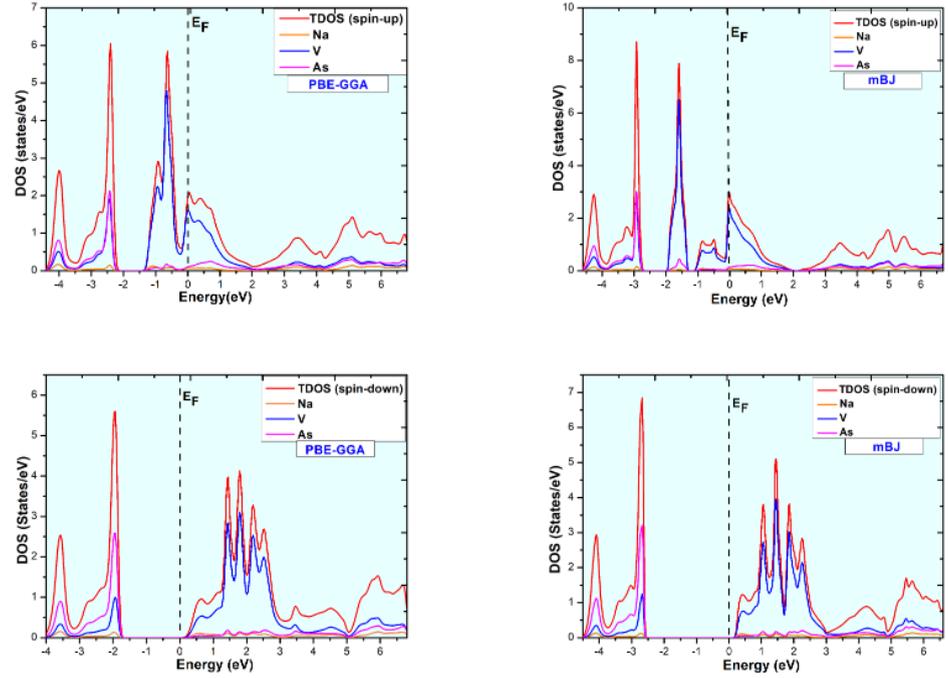

**Figure 4:** Atom resolved PDOS for HH NaVAs calculated using the GGA and TB-mBJ exchange–correlation functionals.

## 4.2. Electrical Conductivity

Electrical conductivity measures how well a material can conduct electric current. It is influenced by factors such as charge carrier concentration, mobility, and temperature, making it critical for assessing a material's thermoelectric efficiency. In computational thermoelectric studies, the ratio of conductivity to relaxation time $(\frac{\sigma}{\tau})$ is evaluated since absolute values of the relaxation time $(\tau)$ are typically cannot be obtained from first-principles calculations. it is given by

$$\frac{\sigma}{\tau} = \frac{ne^2}{m^*} \tag{22}$$

In the above equation n is the carrier concentration, e is the elementary charge, $\mu$ is the mobility, and $m^*$ is the effective mass of carriers. Figure 5(b) illustrates the variation of $(\frac{\sigma}{\tau})$ with respect to the $(\mu - E_F)$ in [Ha] at different temperatures ranging from 600 K to 1500 K for NaVAs. It is clear from figure 5(b) that the $(\frac{\sigma}{\tau})$ profile is asymmetric around the Fermi level. For $(\mu - E_F) > 0$, the ratio of conductivity to relaxation time $(\frac{\sigma}{\tau})$ increases sharply, reaching a value of $(3 \times 10^{20})\ \Omega^{-1}m^{-1}s^{-1}$ with negligible values for $(\mu - E_F) < 0$. This indicates superior contribution from charge carriers of conduction band compared to that from the valence band of NaVAs. The electrical conductivity is found to increase with an increase in temperature near the fermi level which indicates the semiconducting nature of NaVAs. This increase in conductivity contributes positively to both power factor and figure of merit. These results support materials suitability for high performance n-type applications especially in mid to high temperature ranges.

## 4.3. Thermal Conductivity $\kappa$

The electronic thermal conductivity $(\kappa_e)$ is related to $\sigma$ through the Wiedemann-Franz law, which is expressed as:

$$\kappa_e = L\sigma T \tag{23}$$

In the above equation $\kappa_e$ is thermal conductivity, L is Lorentz number, T is absolute temperature and $\sigma$ is electrical conductivity. The thermal conductivity increases with an increase in temperature as per eqn. no. 23. As shown in fig. 5c. the contribution of charge carriers between the Fermi level and the conduction band increases which may be attributed

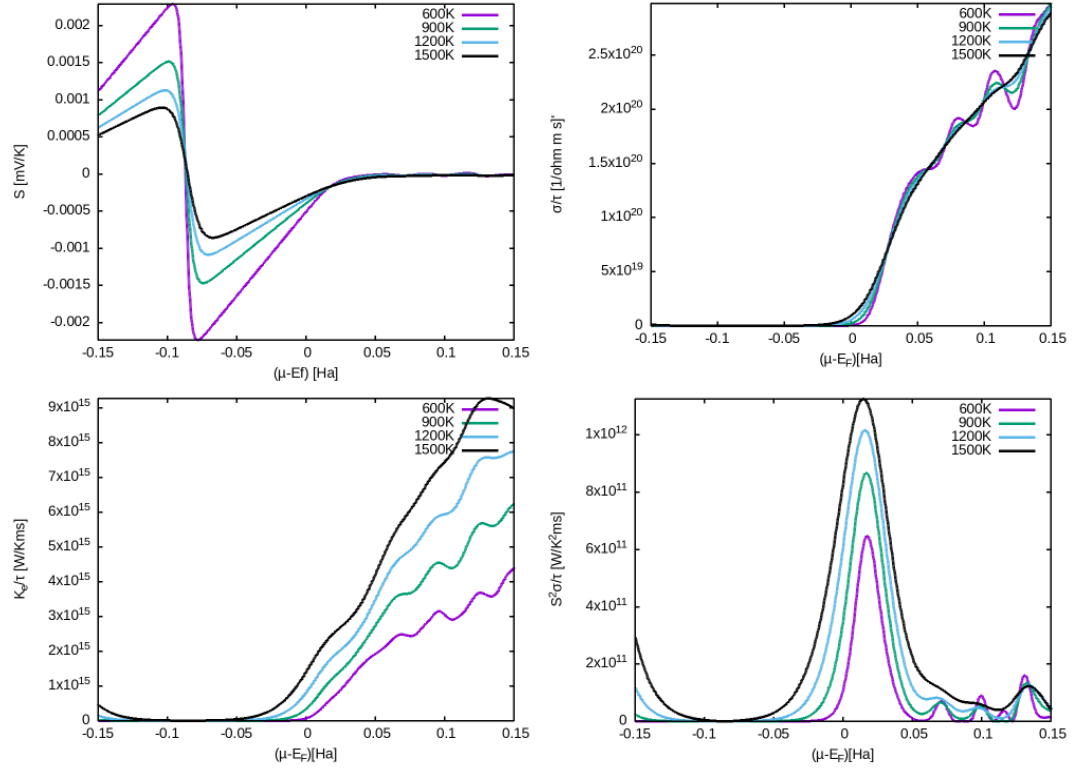

**Figure 5:** Variation of thermoelectric parameters as a function of chemical potential ($\mu - E_F$ [Ha]) at different temperatures [a] Seebeck coefficient [b] Electrical conductivity per relaxation time [c] thermal conductivity per relaxation time [d] Power Factor

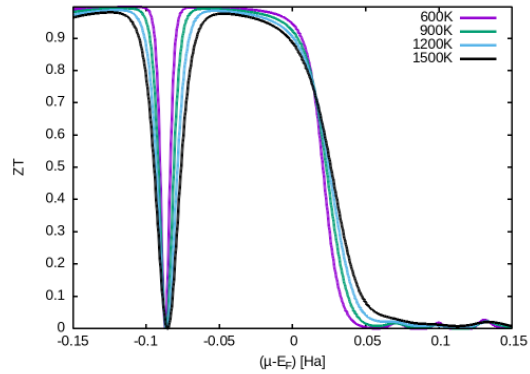

**Figure 6:** Variation of Figure of Merit [ZT] as a function of chemical potential ($\mu - E_F$) [Ha] at different temperatures

to the thermal activation of the charge carriers at higher temperatures.

### 4.4. PowerFactor PF

The power factor (PF), defined as ($\frac{S^2\sigma}{\tau}$), is a crucial parameter for assessing the efficiency of thermoelectric materials. It combines the Seebeck coefficient $S$ and electrical conductivity ($\sigma$) to indicate the effectiveness of a material in converting thermal energy into electrical energy [37]. The power factor per unit relaxation time ($\tau$) is expressed as:

$$PF = \frac{S^2\sigma}{\tau} \tag{24}$$

Figure 5d shows the calculated value of ($\frac{S^2\sigma}{\tau}$) as a function of ( $\mu - E_F$) for the NaVAs alloy at various temperatures, ranging from 600 K to 1500 K. From the fig. 5d we can see that the power factor significantly increases with temperature, reaching its highest values at 1500 K. This trend is consistent with enhanced carrier activation at elevated temperatures, which boosts the electrical conductivity ($\tau$) and optimizes the balance with the Seebeck coefficient (S) in the term ($\frac{S^2\sigma}{\tau}$). The power factor curve displays prominent peaks located at ( $\mu - E_F$) = +0.01 Ha approximately. At 1500 K, the maximum value of ($\frac{S^2\sigma}{\tau}$) is about (1.5 × $10^{12}W/K^2ms$), underscoring the material's excellent potential for thermoelectric applications.

### 4.5. Figure of Merit ZT

The thermoelectric performance of a material is effectively characterized by the dimensionless figure of merit, ( ZT ), as shown in equation 19. A higher ( ZT ) value indicates a better thermoelectric material, which requires large thermopower, high electrical conductivity, and low thermal conductivity [38]. Figure 5e illustrates the variation of the figure of merit ( ZT ) as a function of ( $\mu - E_F$ ) for NaVAs alloy at various temperatures, ranging from 600 K to 1500 K.

Two distinct peaks can be observed in the range of ( $\mu - E_F$ ) = from -0.15 to +0.05 Ha at each temperature. Further, at 600 K, the maximum thermoelectric figure of merit (ZT) approaches $\approx 0.97$ and increases with rising temperature, reaching nearly 1.0 in the range of 900 K to 1500 K. As the temperature increases, the (ZT) peaks decrease in height with dual tone change in its width profile. For energies in the range -0.09 [Ha] < ( $\mu - E_F$ ) < 0.01 [Ha] the peaks shrinks whereas for energies ( $\mu - E_F$ ) greater than 0.01 [Ha] the peaks broaden with temperature This may be attributed to enhanced thermal excitation of charge carriers and enhanced phonon scattering at higher temperatures.

The dip in ZT around $\mu - E_F \approx$ -0.09 [Ha] at all temperatures is due to penetration of fermi level into the conduction band. In summary, we can achieve (ZT) values approaching 1.0 over a broad temperature range. Its performance at high temperatures shows its potential in energy conversion devices, such as thermoelectric generators and waste heat recovery systems.

## 5. Optical properties

The optical properties of a material are essential for understanding how it interacts with electromagnetic radiation. These properties play a crucial role in applications related to optoelectronics. They are derived from the complex dielectric function of the material. To gain a comprehensive understanding of a material's optical behavior, it is important to analyze its frequency-dependent complex dielectric function. This analysis allows us to derive key parameters such as the refractive index, absorption coefficient, reflectivity, and optical conductivity [39].

### 5.1. Absorption Coefficient

The absorption coefficient $\alpha(\omega)$ indicates a material's ability to absorb incoming photons and can be calculated from the dielectric function as:

$$\alpha(\omega) = \frac{\sqrt{2}\omega}{c}\left[\sqrt{\epsilon_1^2(\omega) + \epsilon_2^2(\omega)} - \epsilon_1(\omega)\right]^{\frac{1}{2}} \quad (25)$$

In the above equation $\omega$ represents the angular frequency of the incident photon, while c is the speed of light. The symbols $\epsilon_1(\omega)$ and $\epsilon_2(\omega)$ refer to the real and imaginary parts of the dielectric function, respectively. The absorption coefficient $\alpha(\omega)$ is related to the extinction coefficient $k(\omega)$ through the following equation given as

$$\alpha(\omega) = \frac{4\pi k(\omega)}{\lambda} \quad (26)$$

The Absorption coefficient $k(\omega)$ measures the attenuation of electromagnetic waves as they travel through a material, reflecting energy absorption. As shown in Figure 7a, the Absorption coefficient of NaVAs varies significantly across different energy ranges. A notable peak is observed around 9 eV, indicating strong absorption in this energy range. This peak extend up to approximately 10 eV, suggesting multiple interband transitions. Overall, this behavior implies that NaVAs possess substantial optical absorption capacity across a wide energy spectrum in the visible and ultraviolet (UV) regions.

### 5.2. Real Optical Conductivity

The optical conductivity, denoted as $\sigma_1(\omega)$, is related to the imaginary dielectric function by the following equation [39].

$$\sigma_1(\omega) = \frac{\omega \epsilon_2(\omega)}{4\pi} \quad (27)$$

This quantity indicates the material's ability to conduct electrical current when subjected to an alternating electromagnetic field. The optical conductivity spectrum of NaVAs exhibits values in the range 2.5 eV to nearly 9 eV with a maximum peak near 6.5 eV reaching $\approx 5.5 \times 10^{15}$ per sec. This highlights the presence of strong inter band transitions. The high optical conductivity observed in the visible and (UV) regions.

### 5.3. Optical Reflectivity

Reflectivity $R(\omega)$ measures the fraction of incident light reflected at the surface and is defined by:

$$R(\omega) = \left|\frac{\sqrt{\epsilon(\omega)} - 1}{\sqrt{\epsilon(\omega)} + 1}\right| \quad (28)$$

As illustrated in fig. 7c, the reflectivity of NaVAs shows a nonmonotonic variation with photon energy. Notably, the optical reflectivity remains below 0.5 for photon energies between (2 to 13) eV. This moderate reflectivity behavior across a wide range of energies suggests its potential use in the UV-visible spectrum for reflective coatings. Further the higher values of optical reflectivity in the photon energy range lower than 2 eV suggests its use for infrared shielding.

### 5.4. Energy Loss

The energy loss function $L(\omega)$, defined as the imaginary part of the inverse dielectric function and is given as

$$L(\omega) = Im\left[-\frac{1}{\epsilon(\omega)}\right] \quad (29)$$

It describes the energy lost by fast electrons as they move through a material, which is crucial for understanding plasmonic excitation's. The peak value of L($\omega$) correspond to plasma frequency of that material. As shown in fig. 7d. the plasma frequency for NaVAs will should lie between (12 to 14) eV signifying the dominance of collective oscillations of conduction electrons (plasmons) which is a key characteristic of the metallic nature of Heusler alloys.

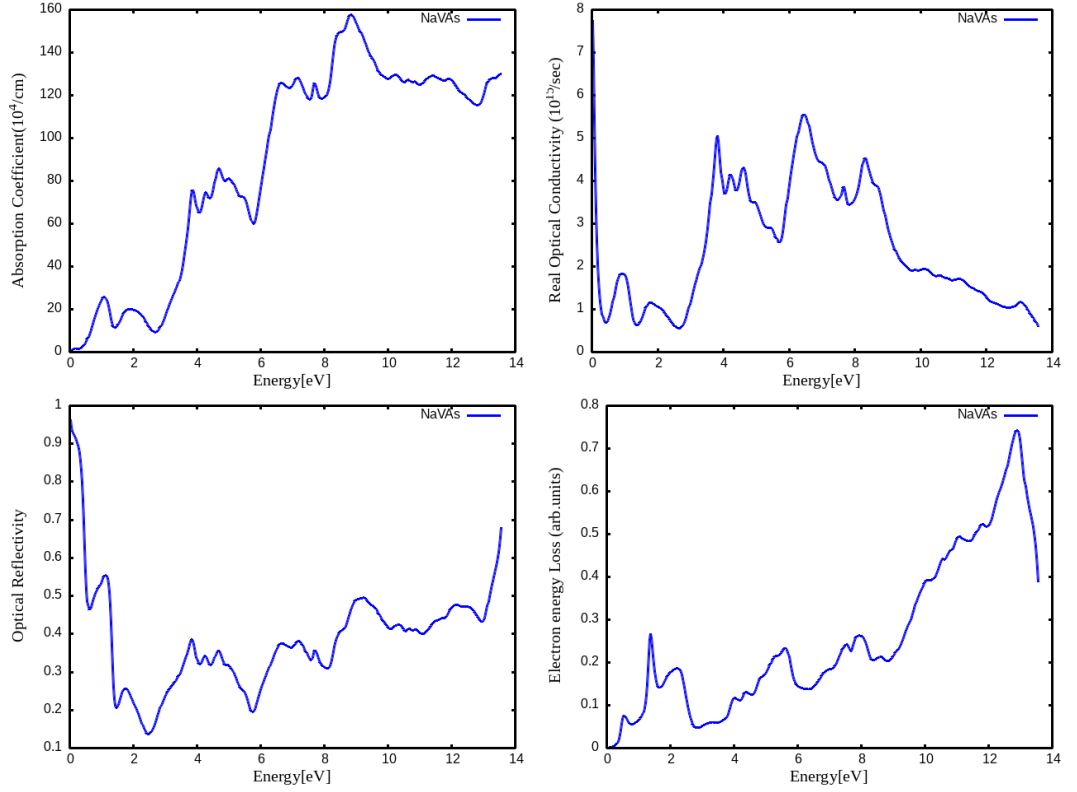

**Figure 7:** Optical properties for NaVAs HH alloy: [a] Absorption Coefficient [b] ] Optical Conductivity, [c] Reflectivity, [d] Energy Loss

## 6. Dynamic Stability

The phonon dispersion curve of NaVAs along the high symmetry path ( X-Γ-L) is shown in Figure 8. This spectrum offers important insights into the dynamical stability and vibrational properties of the material. A significant feature of the phonon dispersion is the absence of imaginary (negative) frequencies throughout the Brillouin zone. This observation confirms the dynamic stability of the NaVAs alloy in its optimized crystal structure. Additionally, the lowest phonon branch, which represents acoustic modes, begins at zero frequency at the (Γ)-point, as is expected due to transnational in variance. A clear phonon band gap is observed between the acoustic and optical modes. Further, some of the modes are found to be degenerate indicating significant inter phonon interactions. These interactions possibly affect the thermal conductivity of the alloy. The high-frequency optical modes play a important role in material's vibrational entropy and there by affect its Debye temperature.

## 7. Thermodynamic Properties

We can ascertain the state of the system and its reaction to external energy and work by examining the thermodynamic response. In reaction to variations in temperature and pressure, this study investigates the alloy's specific heat, thermal expansion coefficient, Debye temperature, and Grüneisen parameter, among other thermodynamic characteristics.Using the quasi-harmonic approximation (QHA) in conjunction with first-principles density functional theory (DFT) calculations, we derive free energy, entropy, heat capacity, and associated parameters over a range of temperatures. Using the GIBBS2 method [40], we used the Quasi-harmonic approximation to determine the thermodynamic properties of the NaVAs alloy at a pressure of 0 GPa and a temperature range of 0 to 1000 K. Understanding the harsh conditions required for thermodynamic applications depends on these computations. The temperature at which a material's constituent vibrations reach their maximum frequencies is known as the Debye temperature ($\theta_D$). The following equation can be used to determine the Debye temperature:

$$\theta_D = \frac{h}{k_B} \left[ \frac{3n}{4\pi} \left( \frac{\rho N_A}{M} \right) \right]^{\frac{1}{3}} V_M \qquad (30)$$

$$V_M = \frac{1}{3} \left[ \frac{2}{V_t^3} + \frac{1}{V_L^3} \right] \qquad (31)$$

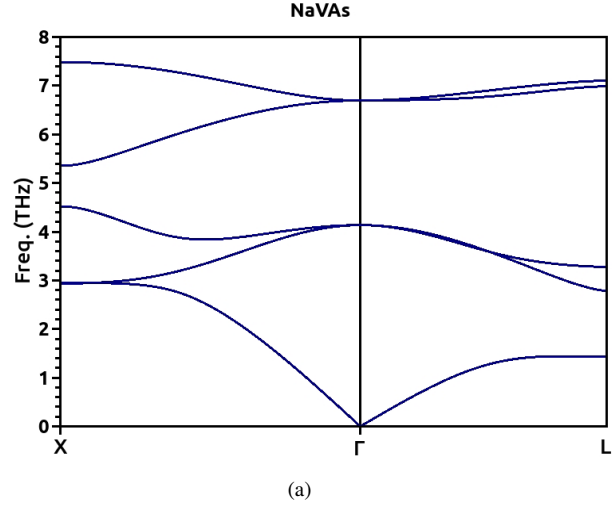

**NaVAs**

(a)

**Figure 8:** Phonon dispersion curve of NaVAs HH alloy

$$V_L = \sqrt{\frac{3B + 4G}{3\rho}} \tag{32}$$

$$V_t = \sqrt{\frac{G}{\rho}} \tag{33}$$

The shear modulus (G) and bulk modulus (B) can be used to calculate specific thermodynamic parameters, such as longitudinal velocity, transverse velocity, and mean velocity of lattice vibrations using equations (30) to (33). According to equation 30, the temperature dependence of the Debye temperature for the NaVAs alloy is illustrated in Figure 9a. It is observed that with a significant increase in temperature, the value of the Debye temperature decreases exponentially. The Debye temperature is a key thermodynamic parameter primarily influenced by lattice dynamics, bonding strength, and the distribution of atomic mass. It measures the highest vibrational frequency of the crystal lattice and is directly related to the speed of sound in the material and the stiffness of chemical bonds. In Figure 9a, we see a notable decrease in the Debye temperature with rising temperature. This phenomenon occurs because, as temperature increases, atomic vibrations become more pronounced, leading to anharmonicity and a reduction in phonon frequencies. Moreover, the change in the Debye temperature is attributed to the strong covalent-metallic bonding between Sodium (Na), Vanadium (V), and Arsenic (As) atoms in the NaVAs system. In crystalline materials, lattice vibrations and phonon excitation are enhanced, resulting in a monotonically increasing entropy for NaVAs with temperature. The observed positive thermal expansion coefficient in the NaVAs system can be explained by the anharmonic lattice interactions that cause the unit cell volume to expand with increasing temperature. Due to increased lattice vibrations and a corresponding rise in phonon population at higher temperatures, the vibrational free energy of the NaVAs half-Heusler alloy increases with temperature, as shown in fig. 9b. When thermal energy is applied, the atoms in the crystal lattice oscillate more vigorously, which raises the vibrational entropy and lowers the Helmholtz free energy. The alloy's inherent phonon softening and anharmonicity are characteristic of thermodynamic stability. To understand lattice thermal conductivity, we consider the Grüneisen parameter ($\gamma$), which measures anharmonicity. Moderate values of the Grüneisen parameter indicate controlled phonon scattering, which is crucial for evaluating thermoelectric efficiency. As shown in fig. 9c, at low temperatures, only low-energy acoustic phonon modes contribute to vibrational free energy (Fvib), resulting in a gradual decrease. As the temperature increases, higher-energy optical modes become activated, causing a more rapid decline in Fvib. This temperature-dependent behavior stabilizes the system by contributing negative free energy, which helps offset the internal energy and lowers the total Gibbs free energy of the alloy. The absence of imaginary phonon frequencies throughout the Brillouin zone confirms the dynamical stability of the system, indicating that the vibrational contributions are physically meaningful and not a result of lattice instabilities. Additionally, the vibrational phonon modes in alkali atoms are relatively soft in NaVAs alloys, leading to higher vibrational entropy contributions that enhance thermal stability at elevated temperatures. The absence of phase transitions or structural distortions within the investigated temperature range (up to approximately 1000 K) is evidenced by the smooth and continuous nature of the vibrational free energy curve. The vibrational free energy supports the thermodynamic stability of the NaVAs alloy, reinforcing its potential for applications in thermoelectric and high-temperature spintronic devices. The vibrational free energy, Fvib, can be expressed as a combination of internal vibrational energy and vibrational entropy. At 0 K, the value of entropy is nearly zero. In fig. 9d, it can be observed that as the temperature rises, internal energy increases, while free energy decreases significantly, in accordance with the thermodynamic equation provided. The alkali metals in the half-Heusler NaVAs contribute minimally to their low-frequency phonon modes,

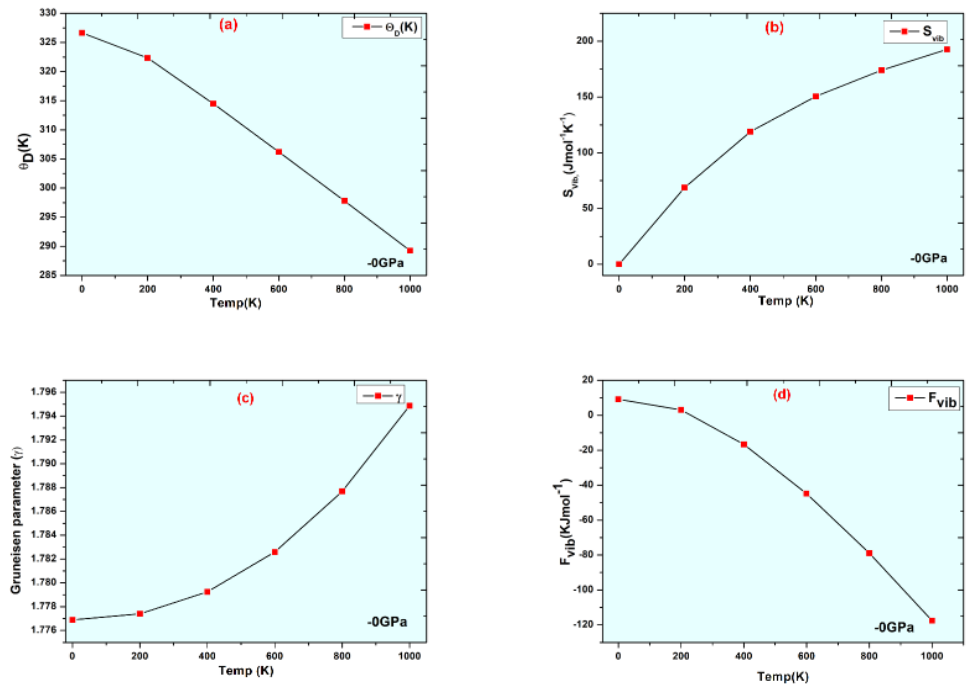

**Figure 9:** Variation of [a] Debye temperature [b] Entropy [c] Gruneisen parameter [d] Vibrational free energy $F_{vib}$ with temperature for NaVAs HH Alloy

which are readily activated with increasing temperature. Consequently, vibrational free energy decreases rapidly with temperature. Another important thermodynamic parameter to consider is heat. We observe in fig.10a that the heat capacity (Cv) is proportional to $T^3$ up to 300 K, after which it reaches saturation and follows the Dulong-Petit law. fig. 10b shows that the vibrational internal energy increases with temperature due to enhanced phonon activity, which results from larger atomic vibrations caused by thermal excitation. In NaVAs, the increased thermal stimulation of lattice vibrations (phonons) leads to a rise in vibrational internal energy as the temperature increases. As the temperature rises, atoms oscillate around their equilibrium positions with greater amplitude, contributing to vibrational energy and populating more phonon modes. This behavior is described by the Bose-Einstein distribution and the phonon density of states, both of which increase with temperature according to the harmonic approximation in density functional theory (DFT). The higher temperature results indicate more atomic vibrations and anharmonic effects within the crystal lattice of NaVAs Heusler alloys. Thus alkali metal-based half-Heusler alloy NaVAs is one of the ternary intermetallic compounds with a cubic crystal structure that shows exceptional thermodynamic behavior, demonstrating its stability and high-temperature application potential.

## 8. Conclusion

In summary, our comprehensive study of the NaVAs half-Heusler alloy highlights its potential as a highly promising candidate for multifunctional applications. Our findings indicate that among the three possible structural configurations, the $\gamma$-type phase is the most energetically stable in the ferromagnetic (FM) phase. Mechanical assessments show that the alloy is stiff, brittle, and auxetic which is supported by calculated mechanical parameters. Phonon dispersion relations reveal no negative frequencies, and the enthalpy of formation confirms the compound's dynamic stability and suitability for experimental synthesis. The electronic structure analysis indicates that NaVAs exhibits half-metallic behavior, with a significant semiconducting band gap of 2.77 eV in the minority spin channel, while maintaining metallic character in the majority spin channel. Density of states (DOS) analysis further corroborates this, demonstrating the contribution of d-block element vanadium (V) near the Fermi level, contributing to its half-metallic nature. The calculated total magnetic moment of 2.99 $\mu_B$ aligns perfectly with the Slater-Pauling rule, confirming the half-metallic ferromagnetic nature of the alloy, with 100% spin polarization at the Fermi level. The thermodynamic properties of the alloy confirm its thermal stability and heat capacity behavior under varying temperature conditions. The observed decreasing trend in Debye temperature with increasing temperature, suggests strong thermal resistance, making it suitable for high-temperature applications. The analysis of heat capacity and entropy trends indicates a stable thermodynamic response. Furthermore, the optical properties demonstrate strong interband transitions and high absorption in the infrared (IR) and ultraviolet (UV) regions, along with characteristics that are beneficial for plasmonic and reflective technologies. Importantly, the thermoelectric study reveals high and temperature-dependent ZT values at with values reaching approximately 1 between 600 K and 1500 K. This emphasizes the potential of NaVAs HH alloy in energy conversion applications. Overall, this work establishes NaVAs as a promising candidate for integration into next-generation spintronics, optoelectronic, and thermoelectric devices.

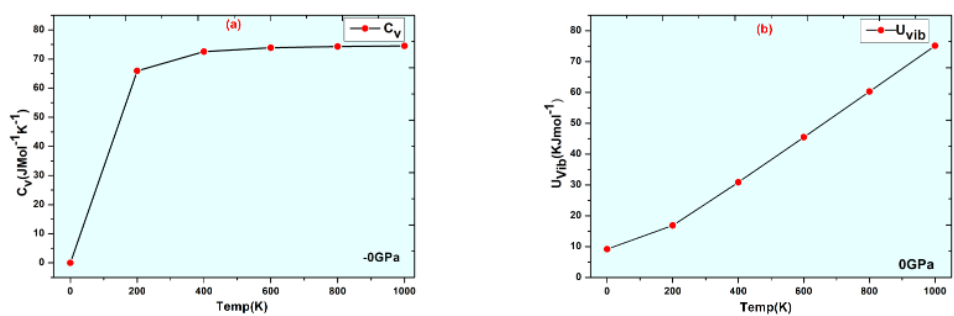

**Figure 10**: Variation of [a] Cv [b] Vibrational internal energy $U_{vib}$ with temperature for NaVAs HH Alloy

## Declaration of Competing Interest

The authors declare that they have no known competing financial interests or personal relationships that could have appeared to influence the work reported in this paper.

## Acknowledgement

Authors gratefully acknowledge the support by Department of Higher Education, Government of Himachal Pradesh, Shimla.

## Data Statement

The data can be provided on request to corresponding author.

## CRediT authorship contribution statement

**Rajinder Singh:** Data compilation, Writing - original draft Writing - review and editing. **Shyam Lal Gupta:** Conceptualization, Methodology, Data curation, Writing - original draft Writing - review and editing. **Sumit Kumar:** Software, Workstation, Data generation. **Lalit Abhilashi:** Software, Workstation, Data generation. **Diwaker:** Conceptualization, Methodology, Data curation, Writing - original draft Writing - review and editing. **Ashwani Kumar:** Data compilation, Writing - original draft Writing - review and editing.